\documentclass[aps,prl,,twocolumn,showpacs]{revtex4}

\usepackage{graphicx}
\usepackage{dcolumn}
\usepackage{bm}

\begin{document}


\title{Efficient excitation of cavity resonances of subwavelength metallic gratings.}


\author{P. Qu\'emerais$^{*}$, A. Barbara , J. Le Perchec and T. L\'opez-R\'ios.}
\email[]{quemerai@grenoble.cnrs.fr}
\affiliation{Laboratoire d'Etudes des Propri\'et\'es Electroniques des
Solides,\\ (LEPES/CNRS), BP 166, 38042 Grenoble Cedex 9, France\\}


\date{\today}

\begin{abstract}
One dimensional rectangular metallic gratings enable enhanced
transmission of light for specific resonance frequencies. Two
kinds of modes participating to enhanced transmission have already
been demonstrated : (i) waveguide modes and (ii) surface plasmon
polaritons (SPP). Since the original paper of Hessel and Oliner \cite{hessel} pointing out the existence of (i), no progress was made in their understanding. We present here a carefull analysis, and show that the coupling between the light and such resonances can be tremendously improved using an {\it evanescent}  wave. This leads to enhanced localisation of light in cavities, yielding, in particular, to a very selective light transmission through these gratings.
\end{abstract}


\maketitle

The practical importance of the optical properties of metallic gratings, is no more to
underline \cite{barnes}. Many important applications were found in
several fields of science, and are today extensively used in many
kinds of sensors and actuators. Recently, a
further property of gratings was found by Ebbesen \textit{et al.}
\cite{ebbesen} : the discovery of an optical transmission
phenomena observed in two-dimensional arrays of subwavelength
holes perforated in a thin metallic plate, which led to
important theoretical
works \cite{porto,moreno,takakura,cao,lalanne}. That experiment is
however in a large extend physically equivalent to a previous one reported
by Dragila \textit{et al.}\cite{dragila}. These authors showed
that a thin metallic opaque plate - without any holes - can
transmit light, provided that it is enlightened by an evanescent
wave generated by total reflection in a prism. For both
experiments, i.e. Dragila \textit{et al.} or Ebbesen \textit{et
al.}, the physics is almost the same. In effect, to excite
a surface plasmon on a flate metallic surface, the incident wave
must have a wavevector component parallel to the surface larger
than $k_0 \sin(\theta)$, $k_0=\omega/c$ being the wavevector of
the incident light, and $\theta$ the incidence angle. There are
two possible ways to do that: by changing the parallel component
of the wavevector into a pseudo-wavevector $k_{//}=k_0 \sin\theta
+ 2 \pi m/d$, where $m$ is a positive or negative integer, and $d$
the period of the holes (case of a perforated plate
\cite{ebbesen}); or by using an attenuated total reflection (ATR)
experiment \cite{otto}. The increase of the parallel
wavevector component is thus provided by the total reflection on a
prism from which an evanescent wave emerges. In
both cases, the surface plasmon polaritons (SPP) of the external
interfaces can be excited. However, another condition is needed to have
transmission through the plate: SPP on both sides must be coupled.
In the case of perforated surface, it is essentially the
evanescent field in the holes which mediates the coupling, while
in the case of ATR, evanescent waves \textit{inside} the metal
directly couples the plasmons of both interfaces. As was pointed
out in
 \cite{barbara1}, for {\it two dimensional} subwavelength holes, the
field in the cavities is always evanescent, so that the coupling, and
thus the efficiency of the transmission, is bounded by the
thickness of the plate: a too thick plate perforated with
subwavelenth holes would transmit almost nothing. It is the same
in the case of ATR experiment where the efficiency of the coupling
is limited by the field decay in the metal, linked to the skin
depth.
\begin{figure}
\includegraphics[scale=0.33]{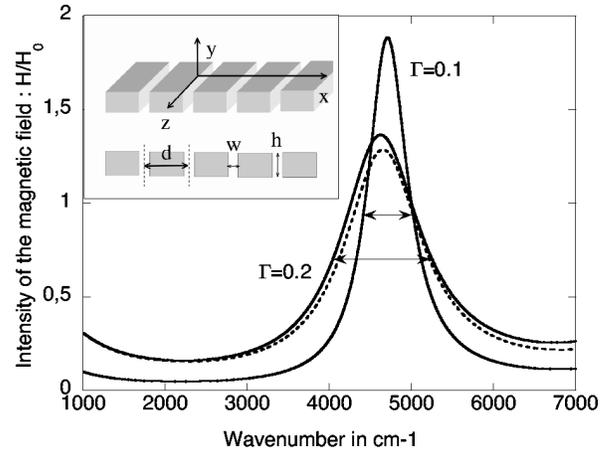}
\caption{Intensity of the magnetic field inside the grooves at a
waveguide resonance for $d=0.5 \mu m$, $h=1 \mu m$, and
$\theta = 0^o$ ($\gamma_0 = 0$, see the text) given for two values of
the geometrical parameter $\Gamma = w/d$ for a perfect conductor
(plain line). A calculation with the dielectric constant of gold (tabulated in \cite{handbook}) for $\Gamma=0.2$ is also represented. }
\end{figure}

Such a thickness limitation may not exist for {\it
one-dimensional} periodically structured film as those sketched on
the inset of fig.1. It was effectively shown experimentally and
theoretically \cite{porto,barbara1,barbara2} that a second
mechanism is important : \textit{the light
can be transmitted by excitation of one of the waveguide modes of
the slits of the gratings} (different than SPPs). Indeed, in this geometry (rectangular
slits), and with a $p$-polarized incident wave i.e. magnetic field
along the $z$-axis, the grating supports modes that propagate into
the slits along the $y$-axis, even when these have a subwavelength
width $w$, and whatever their thickness $h$ is \cite{barbara1}.
However, the transmission by such a mechanism is poorly
selective and the waveguide resonance spectral width is large
\cite{barbara1}. In the infrared region, a half-width of about
$1000$ $cm^{-1}$ was reported for a resonant transmission close to 0.4
occuring at $2530$ $cm^{-1}$ \cite{barbara1}. Recently, using
the same rectangular geometry, but with a more complicated
architecture made of one single slit close to several other
grooves, Martin-Moreno et al. \cite{garcia1} succeeded to
improve the efficiency of the transmission through the slit by
combination of different geometrical parameters which allowed
coincidence of several waveguide resonances.
\begin{figure}
\includegraphics[scale=0.25]{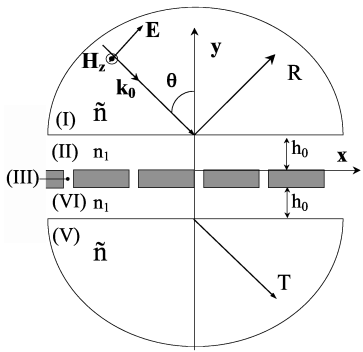}
\caption{ATR configuration using two cylindrical prisms of germanium $\tilde n =4$ surrounding the metallic grating. The angle of incidence satisfies $\theta > \theta_c = \arcsin (1/ \tilde n) \approx 14.5^o$, $n_1$ is taken equal to 1. The zero order transmission is $T=T_0T_0^*$, with $T_0$ given by equ.(\ref{T0}).}
\end{figure}

Waveguide modes of subwavelength metallic gratings in deep
structured surfaces, predicted by Hessel and Oliner
\cite{hessel}, was experimentally observed later \cite{lopez}. The ordinary theory of cavity resonators (see
\cite{jones} for instance) tells us that for any closed metallic
cavity, i.e. a connected volume, there are proper modes which can
carry high electromagnetic field. However, the resonance is
bounded by the absorption of the metallic walls of the cavity
which leads to a finite width in the resonance peaks also called
Q-factor. If the optical absorption is weak, as it is the case for
gold in the infrared, the resonances remain very narrow and
well-defined. Another damping mechanism appears when a hole is
made in the cavity, taking solely its
origin in the geometry. In other words, the \textit{real }
wavevectors of the eigenmodes of a perfectly conducting closed
cavity becomes \textit{complex} by the opening of a hole with a
finite size. This is exactly what happens in deeply structured
gratings where the slits play the role of the hole cited above and
therefore waveguide resonances of gratings (even perfectly conducting) have {\it complex}
wavevectors as was previously pointed out in \cite{hessel}. When they
are excited by a radiating incident wave whose wavevector is {\it
real}, it can not perfectly match with the complex one of the
eigenmode and the resonance is a \textit{forced} one. On the
opposite, we expect a higher Q-factor when exciting the waveguide
resonance with a complex wavevector obtained from an ATR
configuration.

Let us examine these two cases from a simplified theoretical point of
view. We consider a grating with geometrical parameters $d$, $h$
and $w$ (see fig. 1) enlightened (close to the surface) with a
$p$-polarized incident wave $H_z^{(i)}=e^{ik_0 [\gamma_0 x -
\beta_0y]}$. Due to the Helmoltz equation (in the air),
$\gamma_0^2+\beta_0^2=1$. One can choose without any loss of
generality $\gamma_0 \ge 0$. Now, if $\gamma_0 \le 1$, we have a
propagating incident wave with a real wavevector, while if
$\gamma_0
> 1$, we have an evanescent field with complex wavevector. Within the
modal method, used for the calculations and fully described
elsewhere \cite{wirgin,barbara2}, it is easy to show that, in case
of subwavelength slits ($w<< \lambda = 2 \pi /k_0$), the magnetic
field in the slits is proportional to $1 / \Delta(k_0)$, given by
\begin{equation}
\label{delta1}
\Delta (k_0) = (1-D)^2-e^{-2ik_0h}(1+D)^2
\end{equation}
 where $D= \Gamma \sum_{n} S_n^2 / \beta_n$ , $\Gamma = w/d$ is a geometrical factor
of the gratings, $S_n= \sec (k_0 \gamma_n w/2)$ and
$\gamma_n=\gamma_0+n \lambda/d$ and $\beta_n^2=1-\gamma_n^2$ with
$n$ running from $-{\infty}$ to $+{\infty}$. $k_0\gamma_n$ and
$k_0\beta_n$ are thus the wavevector coordinates of the $n^{th}$
reflected order. Waveguide resonances occur at the minima of
$ \mid \Delta \mid^2$. These are approximately reached when $k_0$ satisfies
$\Re (\Delta ) =0$ and the width is then proportional to $\Im (\Delta)$
\cite{wirgin}. Writing $D=\kappa + i\sigma$ and after some
elementary algebra, the equation $\Re(\Delta)=0$ yields to:
\begin{eqnarray}
\label{re=0}
\frac {1} {\tan (k_0 h)} = \frac {(1+ \kappa)^2-\sigma^2} {2 \sigma (1+ \kappa)}.
\end{eqnarray}
which can be easily satisfied by choosing a correct $h$ parameter,
as $\kappa$ and $\sigma$ do not depend on its value.

When equ.(\ref{re=0}) is satisfied, the imaginary part becomes:
$\Im(\Delta)= -2 \kappa$. In order to highlight solely the
resonance phenomena linked to the waveguide modes, {\it we consider all along this paper a
spectral region where no SPP can be excited} ($\lambda
>> d$)\cite{barbara1}. In that case, for any $n\neq0$, all $\beta_n \neq 0$
and are pure imaginary numbers. When the incident wave is propagating, $\gamma_0 \leq 1$ and
$\beta_0$ is real so that $\kappa=\Gamma \sec (k_0 \gamma_0
w/2)^2/\beta_0$. The width $W$ of the resonance has thus a
\textit{finite} value depending on the geometrical parameters of
the grating: $W \propto \Gamma=w/d$. This is shown on figure 1
where the intensity of the magnetic field inside the groove at a
waveguide resonance is represented. This reasoning is made for perfectly conducting gratings.
Including a small absorption in the metal does not quantitatively
change a lot the result as can be seen in figure 1. If, now, $\gamma_0>1$, the incident wave is evanescent since $\beta_0$ becomes {\it imaginary} Consequently,
$\kappa$ and  thus $\Im (\Delta )$ both \textit{vanishe} at the resonance. This is the point we want to
highlight in this letter: an evanescent incident wave would excite much better the
waveguide mode of the grating. 

To illustrate this point, we examine the case of a grating illuminated by
evanescent wave, coming from an ATR configuration as sketched on
figure 2. A first (hemi-cylindrical) prism is positioned above the grating to create
the evanescent incoming wave while the second one plays the
opposite role, converting the evanescent transmitted field into a
propagating one. Both prism have a high refractive index $\tilde n
= 4$ corresponding to that of the Germanium in the infrared
\cite{handbook} and are at a distance $h_0$ from the grating
considered to be a purely gold structure imbedded in air
($n_1=1$). Following the modal expansion method \cite{barbara1,barbara2,wirgin}, the space is
divided into five regions: regions (I) and (V) are the prisms,
(II) and (IV) are the air and (III) is the empty region (air) of the grating, where the
magnetic field is expanded as:
\begin{figure}
\includegraphics[scale=0.33]{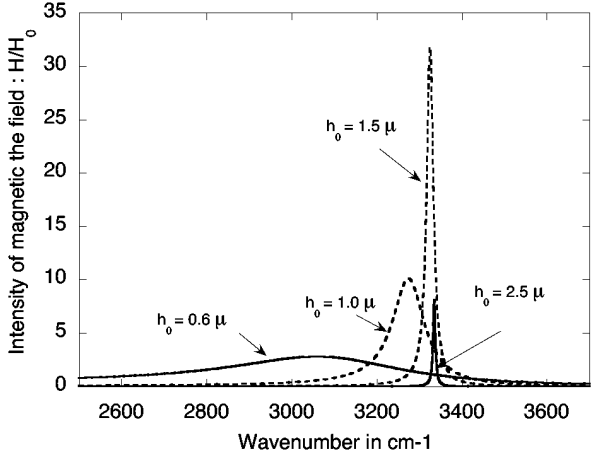}
\caption{The magnetic field intensity for $d=0.5 \mu m$, $h=1 \mu m$, $\Gamma = 0.4$ and $\theta =19^o$, and for different values of $h_0$. The calculations are computed using the device of Fig.2, with $\tilde n=4$, with the optical constants of gold \cite{handbook}. The reduction of the resonance width is clearly visible, as well as the existence of an optimal distance close to $h_0 \approx 1.5 \mu m$.}
\end{figure}
\begin{eqnarray}
\label{field}
{H^{(I)}_{z}(x,y)} &=& e^{ik_{0} \tilde n
[\sin \theta x-\cos \theta (y-h_0)]} \\ \nonumber
\qquad &+& \sum^{+\infty}_{n=-\infty} R_{n} e^{ik_{0} \tilde n
[\gamma_{n}^{(I)} x + \beta_{n}^{(I)} (y-h_0)]}, \\ \nonumber
{H^{(II)}_{z}(x,y)} &=& {\sum^{+\infty}_{n=-\infty}
B_{n} e^{ik_{0} [ \gamma_{n}^{(II)}x-\beta_{n}^{(II)}y]}} \\ \nonumber
\qquad &+& \sum^{+\infty}_{n=-\infty} C_{n}e^{ik_{0}
[\gamma_{n}^{(II)}x+\beta_{n}^{(II)}y]} , \\ \nonumber
{H^{(III)}_{z}(x,y)} &=& A_{0}e^{ik_{0}y}+D_{0}e^{-ik_{0}y},
\\ \nonumber
{H^{(IV)}_{z}(x,y)} &=& {\sum^{+\infty}_{n=-\infty}
B'_{n}e^{ik_{0} [\gamma_{n}^{(IV)}x-\beta_{n}^{(IV)}(y+h)]} } \\ \nonumber
\qquad &+& \sum^{+\infty}_{n=-\infty}  C'_{n}e^{ik_{0}
[\gamma_{n}^{(IV)}x+\beta_{n}^{(IV)}(y+h)]}, \\ \nonumber
{H^{(V)}_{z}(x,y)} &=& \sum^{+\infty}_{n=-\infty} T_{n}e^{ik_{0}
\tilde n [\gamma_{n}^{(V)}x-\beta_{n}^{(V)}(y+h+h_0)]}.
\end{eqnarray}
In each region of space, the field satisfies the Helmholtz
equation, such that whatever $i=$ (I)-(V), and for all $n$,
${\gamma_n^{(i)}}^2+ {\beta_n^{(i)}}^2=1$. Moreover,
$\gamma_n^{(I)}=\gamma_n^{(V)}= \sin \theta + n \lambda / \tilde n d$, and
$\gamma_n^{(II)}=\gamma_n^{(IV)}= \tilde n \gamma_n^{(I)}$. The
expression of the field in (III) is given only taking into account
the first term $n=0$ of a more general expansion, which is a very good approximation in the case where
$\lambda >>w$ \cite{barbara2}. The amplitudes of the field in the
grooves $A_0$ and $D_0$ as well as the reflection and transmission
coefficient are calculated by applying the usual boundary
conditions \cite{barbara1} :
\begin{eqnarray}
\label{a0d0} {A_0} &=& V_0(1-D^{+}) /  \tilde \Delta(k_0)
\nonumber \\ \nonumber {D_0} &=& -V_0 e^{-2ik_{0}h} (1+D^{-}) /
\tilde \Delta(k_0),
\end{eqnarray}
with
\begin{eqnarray}
\nonumber \tilde \Delta(k_0) &= &
(1-D^{+})^2-e^{-2ik_{0}h}(1+D^{-})^2  \\ \nonumber \\ \nonumber
D^{\pm} &=& \Gamma \sum_{n=- \infty}^{+ \infty} S_n^2 {\frac {b_n
(1\pm \eta)} {\eta b_n + \beta_n^{(II)} d_n}}  \\ \nonumber
\\\nonumber V_0 &=&  \bigg( \frac { 2 X_0 \beta^{(II)}_0 S_0} { b_0 \eta + d_0 \beta_0^{(II)}} \bigg).
\end{eqnarray}
with $\eta = 1/ \sqrt{\varepsilon}$ where $\varepsilon$ is the
metal dielectric constant (the limit for a perfectly conducting
material is obtained by taking $\eta \to 0$). The terms $b_n$ and $d_n$ are expressions given by:
\begin{eqnarray}
b_n&=&\cos Z_n(1-iX_n \tan Z_n) \nonumber \\
d_n&=&\cos Z_n(X_n-i \tan Z_n) \nonumber
\end{eqnarray}
where $Z_n=k_0\beta^{(II)}_n h_0$ and
$X_n={\beta^{(I)}_n} / \tilde n {\beta^{(II)}_n}$.
\begin{figure}
\includegraphics[scale=0.33]{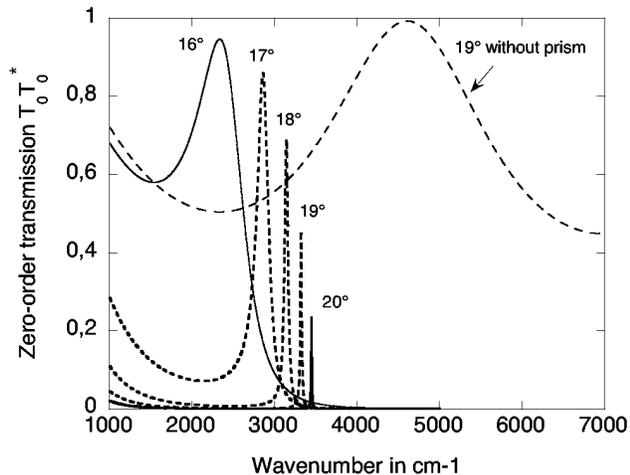}
\caption{Absolute zero-order transmission for different angle of incidence $\theta > \theta_c = 14.5^o$ for $h_0=1.5$ $\mu m$. The calculations are computed for the device of Fig.2 with $\tilde n =4$ and the same grating as in Fig.3. By comparaison, the zero-order transmission of the same gratings without prism is also represented for $\theta = 19^o$. With prism, the resonance width is about $20$ $ cm^{-1}$, while without prism, it is about $1500$ $cm^{-1}$.}
\end{figure}
The zero-order transmission coefficient is then:
\begin{equation}
\label{T0}
\nonumber T_0 =- \Gamma S_0 \Bigg[ \frac {(1-\eta)
e^{-ik_0h}A_0-(1+ \eta) e^{ik_0h}D_0}{ \eta b_0 + d_0
\beta_0^{(II)}} \Bigg].
\end{equation}

To have a total reflexion at the interface (I)/(II), we consider an angle of incidence which satisfies $\theta > \theta_c = \arcsin (1/ \tilde n)$. Under such condition, all the different orders of the field in region (II) and (IV) are evanescent since $\beta_n^{(II)}=\beta_n^{(IV)}$ are pure imaginary numbers whatever $n$ is. Let us first examine the assymptotic case $h_0 \to + \infty$, which clarifies the general trend of the system. Just at the resonance, one can show after some algebra (linear expansion in $\eta$ which is small) that for large enough $ \delta_0= \mid Z_0 \mid$, the modulus of $\tilde \Delta(k_0)$ reduces to $\mid \tilde \Delta \mid \sim  O( e^{-2 \delta_0 })+ O( \mid \eta \mid)$, where $O(x)$ means "quantity of order of $x$". Subsequently, the field in the groove behaves as : $ \mid H^{(III)} \mid \sim O( e^{- \delta_0} ) /  [ O( e^{-2 \delta_0})+ O( \mid \eta \mid) ]$, while the zero order transmission coefficient $\mid T_0 \mid \sim O( e^{- 2 \delta_0} ) /  [ O( e^{-2 \delta_0})+ O( \mid \eta \mid) ]$. By noting the difference of factor in the exponential terms ($e^{-2 \delta_0}$  and $e^{- \delta_0}$), one immediately sees that for a perfect metal, i.e $\eta = 0$ strictly, the width of the resonance tends to zero as $h_0 \to + \infty$ (recall that $ \delta_0 \propto h_0$). In the same time, the field in the groove exponentially increases, while the transmission tends to 1 just at the resonance frequency and to zero elsewhere. For a real metal, i.e with finite (even very small) $\eta$, there is a qualitative change. First, the width of the resonance is bounded, and tends to a very small, but finite value $O( \mid \eta \mid)$. In that case both the field and the transmission vanish owing to the exponential factor in the numerator. All these statements, can be understood physically : as $h_0$ becomes larger, the number of photons which arrives on the grating exponentially vanishes. If there is no absorption at all, the resonance can be excited, while if there is an absorption, a part of incident photons are absorbed, while the remaining ones excite the resonance. When there is no photons anymore, the resonance cannot be excited. This effect is shown on fig.3, where the resonance width is strongly reduced as expected and a maximum of enhancement of the field is observed around $h_0 = 1.5$ $\mu m$. For larger $h_0$ the field decreases as can be seen for $h_0=2.5$ $\mu m$. The same phenomena arises for the transmission which is directly connected to the field enhancement in the slits. A compromise is thus necessary between the gain in the width of the resonance, which always decreases with increasing $h_0$ up to its lower bound $O( \mid \eta \mid)$), and the efficiency of the transmission. The angle of incidence also enters in the expression of $Z_0$ so that the results are very sensitive to it owing to the exponential dependance of the different terms. Fig.4 shows the zero order transmission for $h_0=1.5$ $\mu m$ for different angles. As can be seen, the reduction of the resonance width $W$ of the transmission is spectacular. The resonance widths obtained for $\theta = 17^o$, $18^o$, $19^o$, and $20^o$, are respectively $W=170$ $cm^{-1}$, $50$ $cm^{-1}$, $20$ $cm^{-1}$ and $9$ $cm^{-1}$, while the transmissions are maintened to high values, respectively $0.85$, $0.70$, $0.45$, and $0.25$. These kind of devices could  be employed to create polarization filters with extremely narrow widths.
Finally, we have shown in this paper that the cavity resonances excitation can be notably improved, yielding a strong localization of the electromagnetic field. The key point is the use of an evanescent incident wave, illustrated here by an ATR configuration. However, many other devices producing evanescent waves could also be employed to enhance the field in such cavities.

\end{document}